\documentclass[aps,twocolumn,amsmath,amssymb,amsfonts,floatfix,superscriptaddress,showkeys,showpacs]{revtex4-1}
\usepackage{dcolumn}  % Align table columns on decimal point
\usepackage{multirow}
\usepackage[T1]{fontenc}
\usepackage[latin1]{inputenc}
\usepackage{dsfont}
\usepackage{verbatim}
\usepackage{bm} % bold math 
\usepackage{hyperref} 
\usepackage{cleveref} 
\usepackage{floatrow}
\usepackage{natbib}
\usepackage{graphicx}
\usepackage{epstopdf}
\usepackage{ifpdf} % This lets us get figures right both in pdf and ps versions
\usepackage{epsfig}
\usepackage{color}
\usepackage[usenames,dvipsnames]{xcolor}
\usepackage[caption=false]{subfig}
\usepackage[export]{adjustbox}
\usepackage[colorinlistoftodos]{todonotes}
\begin{document}

\title{Re-entrant bimodality in spheroidal chiral  swimmers in shear flow}

\author{Hossein Nili}
\email[Corresponding author: ]{hnili@ipm.ir}
\affiliation{School of Physics, Institute for Research in Fundamental Sciences (IPM), Tehran 19395-5531, Iran}

\author{Ali Naji}
\email[]{a.naji@ipm.ir}
\affiliation{School of Physics, Institute for Research in Fundamental Sciences (IPM), Tehran 19395-5531, Iran}

%\date{\today}

\begin{abstract}
We use a continuum model to report on the behavior of a dilute suspension of chiral swimmers subject to externally imposed shear in a planar channel. Swimmer orientation in response to the imposed shear can be characterized by two distinct phases of behavior, corresponding to unimodal or bimodal distribution functions for swimmer orientation along the channel. These phases indicate the occurrence (or not) of a population splitting phenomenon changing the swimming direction of a macroscopic fraction of active particles to the exact opposite of that dictated by the imposed flow. We present a detailed quantitative analysis elucidating the complexities added to the population splitting behavior of swimmers when they are chiral. In particular, the transition from unimodal to bimodal and vice versa are shown to display a re-entrant behavior across the parameter space spanned by varying the chiral angular speed. We also present the notable effects of particle aspect ratio and self-propulsion speed on system phase behavior and discuss potential implications of our results in applications such as swimmer separation/sorting.  
\end{abstract}

\maketitle

\section{Introduction}

Self-propelled  micro-/nano-swimmers have garnered increased interest over the past few decades \cite{marchetti-review,ebbenshowse-review,abp-review,sntln-shelley-review}. One important motive has been the abundance of self-propelled particles in nature. This includes the vast majority of bacteria \cite{bergbook,bacteria-lauga}, sperm cells \cite{sperm-woolley,sperm-alvarez-review}, and algae \cite{green-algae-goldstein,algae-drescher}. Inspired by nature, many artificial swimmers have been realized that swim in fluid environments using different mechanisms \cite{artificial-sperm,ramin-colloidal-prl,paxton}. Among other (bio)technological applications, artificial swimmers bring the prospect of drug delivery on the nano-scale \cite{drugdelivery-lauga}. Biological or artificial, the motion of small-scale swimmers in fluid media is governed by low-Reynolds-number hydrodynamics \cite{purcell}, given the small sizes and low self-propulsion speeds that are typical of these particles. 

Swimmers are most commonly found in confined environments (e.g., microfluidic channels or physiological pathways), and are subject to a form of shear. The self-propulsion of biological swimmers is an inherent feature that helps them follow or avoid the different forms of external stimuli (e.g., nutrients, chemical toxins, light, etc.) in the environment. Given the dynamic nature of fluid environments (especially biological/physiological), the swimmers would have to change orientation every while to adapt to, and optimally survive in, the changing environment. This pattern of motion is known as run-and-tumble \cite{runtumble-elgetigompper,runtumble-ramin}, and is complemented by translational and rotational diffusion of the active particles. The presence of external shear in the environment only complicates the strategy that would need to be adopted by the swimmers in their motion \cite{stocker-natrev,ecoli-upstream,ecoli-jefferey}. In confined environments, active particles show a propensity to move toward and accumulate on confining boundaries (e.g., channel walls) \cite{sperm-woolley,near-wall-classic,berke-near-wall}. This tendency of swimmers has led to focussed interest on the near-wall behavior of active particles \cite{ardekani-nearwall,brown-uni_near-wall,Mathijssen2015,zoettl}. 

The mechanics of swimmer self-propulsion in fluid environments can be described using the notion of a force dipole: Two equal and opposite forces, on (by) the fluid by (on) the particle \cite{batchelor1970}. The vast majority of biological swimmers are asymmetrically shaped, exhibiting a form of chirality. Artificial swimmers, too, commonly exhibit chirality, due to fabrication inaccuracies, or indeed by design, for different tasks and purposes desired of the particles \cite{ghosh-fischer,yeomans-lowen-circle-swimmer,chiral-clusters}. The asymmetry leads to misalignment between the line of self-propulsion and the force dipole, hence a torque experienced by chiral swimmers that works as an extra factor (alongside shear and rotational diffusion) affecting swimmer orientation. The chiral geometry gives rise to hydrodynamic coupling between translational and rotational motion of the low-Reynolds swimmers \cite{kraft2013}, with the active particles rotating simultaneous with their translational motion. The repeated rotation and translation patterns of motion result in chiral swimmers following helical trajectories in three dimensions (3D), and circular trajectories in two dimensions (2D) \cite{lowen-review-chiral}. While the circular (2D) swimming of micro-organisms was realized and studied from long ago \cite{jennings1901}, it was thanks to the development of advanced 3D tracking and imaging techniques that the true 3D swimming of biological swimmers was brought to light \cite{crenshaw1996}. The chemotaxis of biological swimmers (such as sperm cells) toward attractants in the fluid environment is also known to follow helical paths \cite{julicher-prl-chiral,sperm-chiral-ribbon}. %Helical trajectories have been observed and analysed for a variety of micro-organisms including sperm cells \cite{sperm-chiral-ribbon} and bacteria. 
As swimmer geometry is key to the helical pattern of motion, artificial swimmers have also been designed to follow helical paths. As an example (among many), biomimetic bacteria that use artificial flagella have been shown to follow helical trajectories in two directions, with the swimming induced by particle chirality \cite{artificial-flagella}. Even camphors, with a nearly spherical geometry, have been shown to feature helical swimming \cite{camphor}. With swimmer geometry crucial to the performance of chiral active particles, the structures of artificial swimmers can be optimized for best performance, e.g., significantly increased  self-propulsion speeds \cite{chiral-geom-opt}.

In this work, we study the steady-state behavior of a dilute suspension of chiral swimmers confined by the walls of a planar channel and subject to externally imposed shear with a linear profile across the channel width (Couette flow). We model the chiral swimmers as spheroidal particles of varying aspect ratio, and report on the effect of particle chirality and aspect ratio on their overall swimming behavior. Given the importance of the near-wall behavior of active particles, we choose the top wall of the channel (with no loss of generality) as test region to display our results. We specifically report on how the population splitting of active particles into distinct oppositely swimming (downstream and upstream) sub-populations, arising -- in the case of non-chiral swimmers \cite{popsplit-paper} -- from imposed shear rate surpassing a threshold, is altered qualitatively when the swimmers are chiral and exhibit finite thickness. 

\section{Model and continuum method}
\label{sec:model}

\begin{figure}[t!]
\sidesubfloat[]{%
  \label{channel}
  \includegraphics[width=0.85\linewidth]{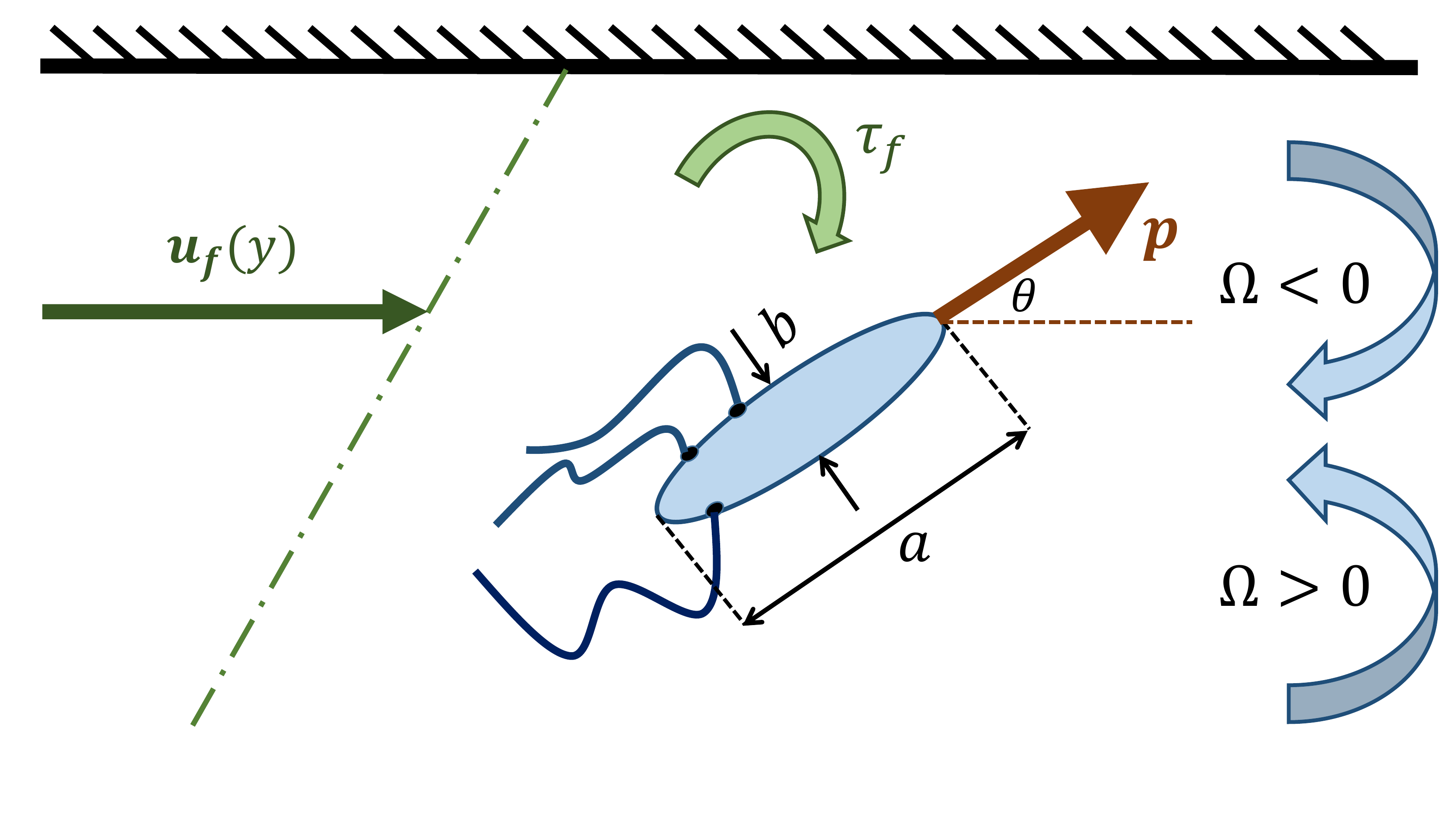}%
}\\
\sidesubfloat[]{%
  \label{torques}
  \includegraphics[width=0.85\linewidth]{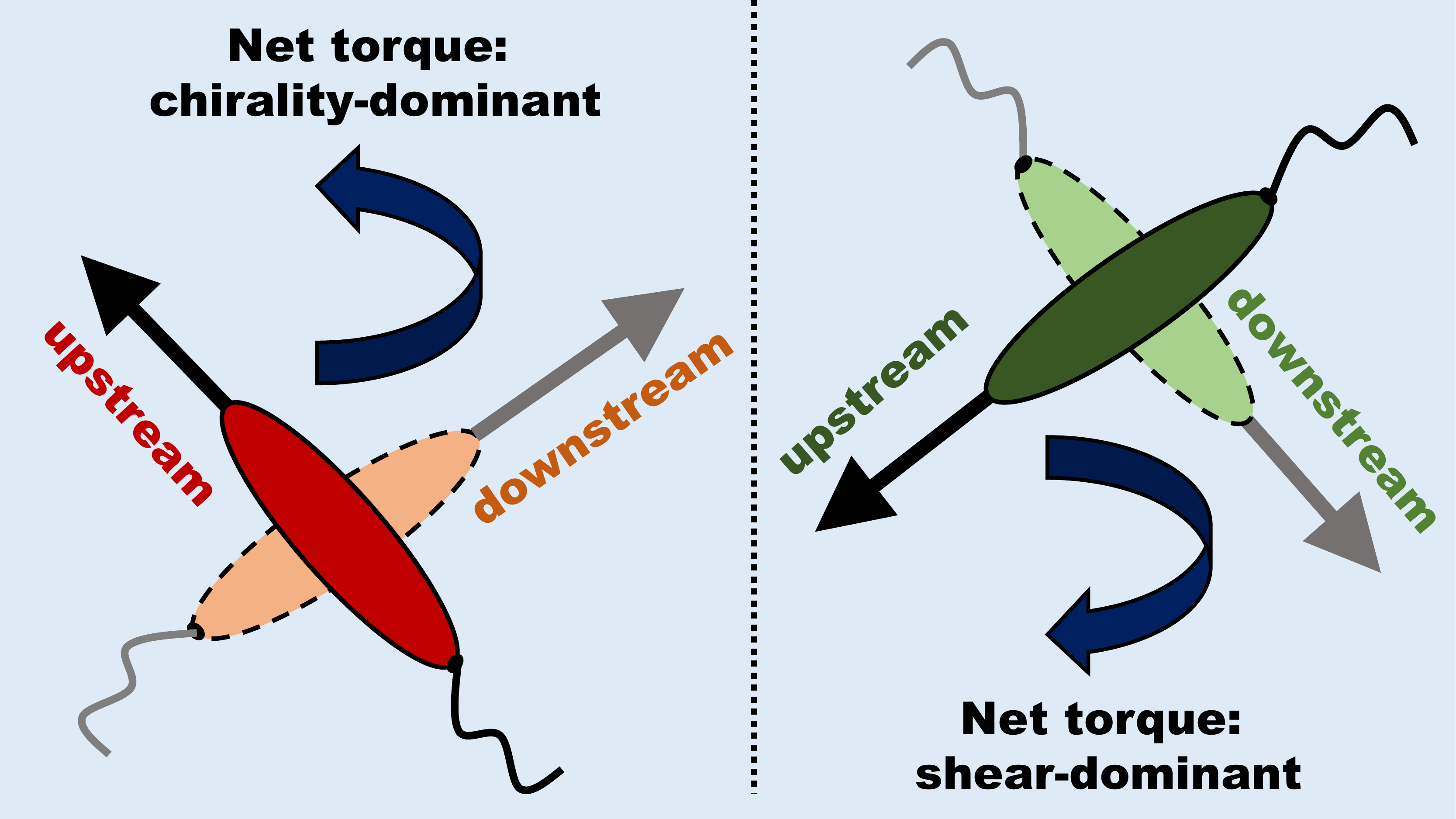}%
}
\caption{(a) Sample spheroidal self-propelled particle with major and minor axes of lengths $a$ and $b$ swimming downstream (i.e., with an orientation vector $\mathbf{p}$ making an angle $-\pi/2<\theta<\pi/2$ with the positive horizontal axis) near the top wall of a channel subjected to an imposed Couette flow. The swimmers are chiral, with angular speeds $\Omega$ with both levogyre and dextrogyre chiralities ($\Omega>0$ or $<0$, respectively) permitted. The torque from flow (always clockwise in the current settings) acting on the active particles is shown as $\tau_{f}$; (b) For a given imposed shear, there are two angular speeds at which a fraction of downstream-swimming (majority) chiral particles flip their swimming direction to upstream, leading to the emergence of a minority population. At a smaller angular speed (right schematic), the conversion is dominated by imposed shear, and at a larger angular speed (left schematic), chirality overtakes the effect of imposed shear, and leads to a second population splitting, marking re-entrant bimodality of the active suspension.
}
\label{schematic}
\end{figure}

Physical specifications of the system we study is shown in Fig. \ref{channel}. We consider an active suspension of chiral self-propelled particles that we model as spheroids with major and minor axes of lengths $a$ and $b$, respectively, giving aspect ratio $\lambda=a/b$. The swimmer orientation, denoting its active self-propulsion, is represented by orientation vector $\mathbf{p}$ that makes an angle $\theta$ with the positive horizontal axis. Both levogyre and dextrogyre chirality are schematically depicted, corresponding to counter-clockwise (CCW) and clockwise (CW) rotations of the swimmers, or positive and negative angular speed $\Omega$, respectively. 

The active suspension is confined by the walls of a channel of half-width $H$, and an external flow is imposed onto the system that we assume to have a linear profile $\mathbf{u}_f(y)$, namely a Couette flow, directed along the horizontal axis $x$ with shear rate $\dot{\gamma}={\partial \mathbf{u}_f(y)}/{\partial y}=U_{m}/2H$, where $y$ is the direction perpendicular to the flow, and $U_{m}$ is the maximum flow velocity, at which we could assume the top wall (at $y=+H$) is moved, while the bottom wall (at $y=-H$) remains stationary. With the given structure of imposed flow, the torque $\tau_{f}$ it exerts on the chiral swimmers will always act to move them in the clockwise (CW) direction. As such, levogyre chirality acts in opposition and dextrogyre chirality in concert with the flow in the torque they exert on the active particles; this can be seen from the schematic of Fig. \ref{schematic}.  

%%%FIG%%%%%%%%%%%%%%%%%%%%%
\floatsetup[figure]{style=plain,subcapbesideposition=top}
%%%%%%%%%%%%%%%%%%%%%%%%%

We adopt a continuum model of swimmer behavior that has been presented and discussed in a number of studies \cite{saintillan-pof,ezhilan}. The model is based on the Smoluchowski equation, expressing conservation of swimmer numbers, and solves for the probability distribution function (PDF), $\Psi(y,\theta)$. We look at the steady-state behavior of the active suspension, hence the absence of time in the independent variables. Also, the symmetry of the problem implies $x$-independence. Chirality of the particles, as well as their geometry (finite aspect ratio), affects the rotational flux velocity of the swimmers, $\dot{\theta}=\dot{\gamma}(\beta \cos(2\theta)-1)/2$, where $\beta$ is the Bretherton shape parameter \cite{bretherton} given, in terms of particle aspect ratio $\gamma$, as $\beta=(\lambda^2-1)/(\lambda^2+1)$. From this, the Smoluchowski equation governing system behavior takes the following form, after the non-dimensionalization of the vertical coordinate with the channel half-width as $y\rightarrow y/H$:
\begin{align} \label{smol-nond}
\xi^{2}\frac{\partial^{2}\Psi}{\partial y^{2}}+\frac{\partial^{2}\Psi}{\partial \theta^{2}} &=\frac{\partial}{\partial \theta}\left\lbrace\Psi\left[\frac{1}{2}Pe_{f}(\beta \cos(2\theta)-1)+\Gamma\right]\right\rbrace\nonumber\\
&+\frac{\partial}{\partial y}(2Pe_{s}\Psi \sin\theta), 
\end{align}
where the $\Gamma=\Omega/D_R$ is the dimensionless angular speed of the chiral swimmers, and $Pe_s=V_s/(2HD_R)$ and  $Pe_f=U_m/(2HD_R)$  are the swim and flow P\'eclet numbers, representing the (relative) strengths of active self-propulsion and imposed shear, respectively. The third dimensionless parameter in Eq. \eqref{smol-nond}, $\xi^{2}=D_{t}/(D_RH^{2})$, serves as a measure of  channel confinement of swimmers. Also, $D_{t}$ and $D_R$ are translational and rotational diffusion coefficients of  the spheroidal swimmer, for which we use the expressions derived by Koenig \cite{koenig} that provided corrections to original formulae by Perrin \cite{perrin}. 

All calculations, for numerical solution of the governing Eq. \eqref{smol-nond}, were done in COMSOL Multiphysics v5.2a; our previous work \cite{popsplit-paper} contains the details. On top of the dimensionless parameters defined under Eq. \eqref{smol-nond}, we use the following as simulation parameters (all derivable from the three dimensionless numbers of the governing equation): $n_s$, giving the number of particle lengths (major axis) that the particle swims in a unit of time; $n_H$, the ratio of channel half-width $H$ over particle length (major axis); $n_F$, the ratio of maximum imposed flow speed (at top channel wall), $U_{m}$, over the swimmer self-propulsion speed $V_{s}$.     

\section{Results and Discussion}

\subsection{Specifications of the baseline parameters} 
\label{baseline}

Our baseline (used as reference) active suspension comprises of prolate spheroidal particles with aspect ratio $\lambda=4$, corresponding to the aspect ratio an \textit{E. coli} bacterium with $a=2\,\mu {\mathrm{m}}$, $b=0.5\,\mu {\mathrm{m}}$. To observe the effect of particle aspect ratio on system behavior, we shall keep the major axis length fixed at $a=2\,\mu {\mathrm{m}}$, and will vary $b$ over a wide range, covering particles close in shape to a sphere to those that are needle-shaped. We should like to stress that our goal is to analyze the generic behavior of spheroidal chiral 
%we do not intend for this work to be adapted to \textit{all} specifications of \textit{E. coli} cells, as we aim
swimmers over a wide range of values for the system parameters introduced in Section \ref{sec:model}, rather than any particular swimmer-specific features. 
%  such as  particle aspect ratio, particle angular speed, swimmer self-propulsion speed, and applied shear rate  
Since we base our analysis on a dimensionless representation, the results reported for a fixed set of dimensionless parameters will be applicable to any set of actual (or dimensional) parameter values (such as channel width, swimmer semi-axis dimensions, self-propulsion velocity, etc.) as long as they can be mapped to the same values of the dimensionless parameters. However, for the sake of concreteness, and extending on the  default parameter values, we make the following as baseline, so that the actual parameters will have these values unless otherwise stated: $V_s=2\,\mu {\mathrm{m}}/{\mathrm{s}}$ and $U_{m}=200\,\mu {\mathrm{m}}/{\mathrm{s}}$, corresponding to self-propulsion and shear factors $n_{s}=0.5$ and $n_F=100$, respectively; and $n_{H}=5$, giving the channel a width of $2H=2n_{H}a=
%2(5)(2\,\mu {\mathrm{m}})=
20\,\mu {\mathrm{m}}$. The translational and rotational diffusion coefficients, derived from the parameter values, are $D_t=2.3\times 10^{-13}$ $m^{2}/{\mathrm{s}}$, and $D_R=0.2/{\mathrm{s}}$.  The dimensionless parameters will have the following values: $Pe_s=0.25$, $Pe_f=25$, and $\xi=0.1$. 

\subsection{Effect of chirality on swimmer distribution} 

\begin{figure}
	\centering
	\includegraphics[width=0.95\linewidth]{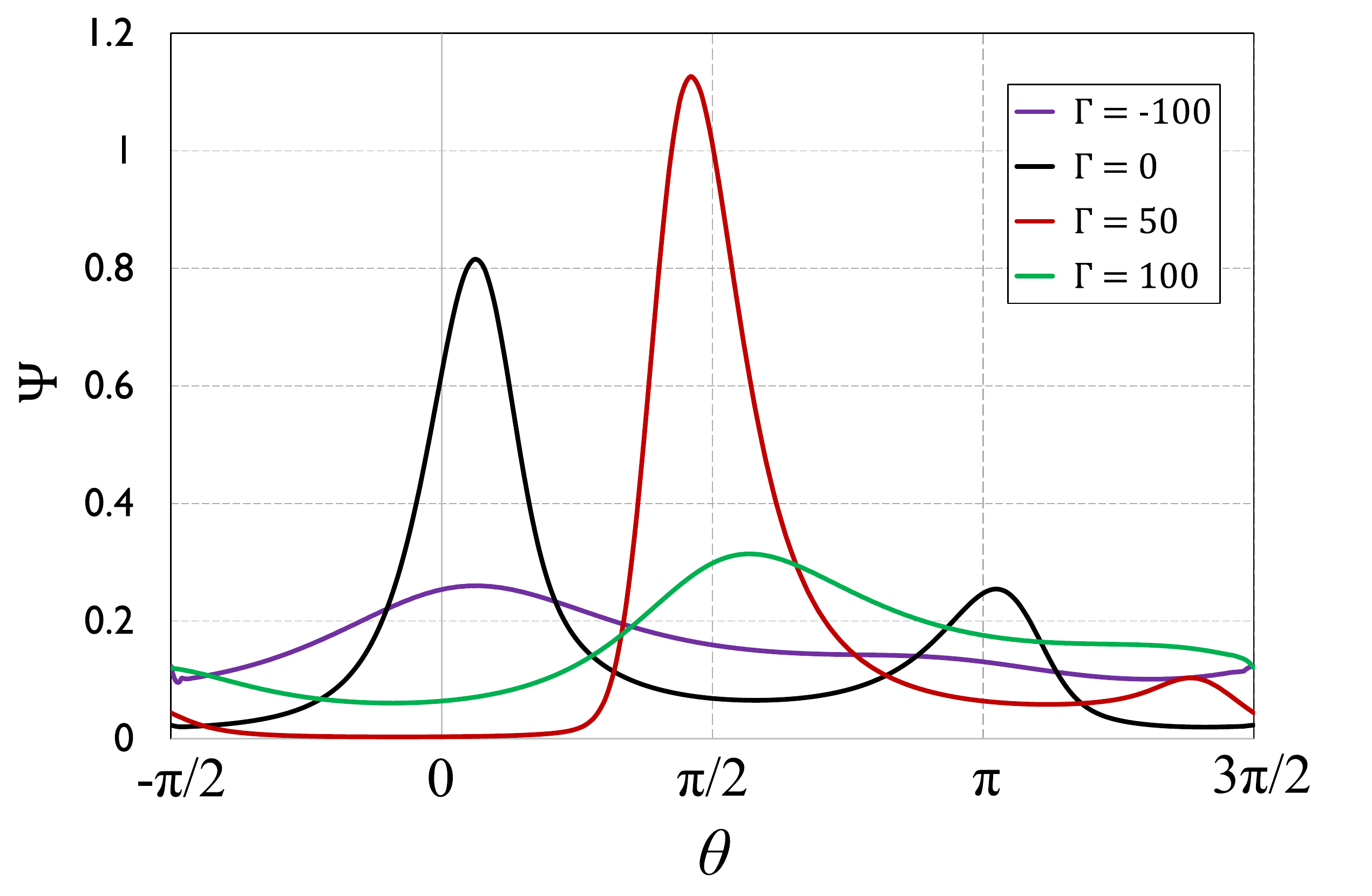}
	\caption{Rescaled orientational  PDF (normalized by total concentration in the channel) of spheroidal chiral swimmers of aspect ratio $\lambda=4$ (corresponding, e.g., to $a=2\,\mu {\mathrm{m}}$ and $b=0.5\,\mu {\mathrm{m}}$) on the top wall of the channel, when they exhibit angular speeds of different magnitudes and directions, including the non-chiral case ($\Gamma=0$).\label{distribution}}
\end{figure} 

For the baseline set of parameter values, Fig. \ref{distribution} shows the re-scaled swimmer PDF across the whole $(0,2\pi)$ range of swimmer orientation angles $\theta$. The horizontal axis ($\theta$) has been set to start from $\theta=-\pi/2$ and end at $\theta=3\pi/2$, so that (for clearer display) the first and second halves of the axis correspond to active particles swimming {\em downstream} and {\em upstream}, respectively. The plot shows the effect of angular speed $\Gamma$ of the chiral active particles on swimmer distribution, where (given the importance of near-wall swimmer behavior) we have chosen the top wall of the channel as test region to display our results. The case of non-chiral swimmers ($\Gamma=0$) has been shown for comparison. It can be seen that were the swimmers non-chiral, there will be splitting of swimmer population into majority (downstream) and minority (upstream) sub-populations, represented by the two peaks in the swimmer PDF. This shows that with no chirality in the picture, the imposed flow has sufficient strength to overturn the swimming direction of a sizeable fraction of active particles from downstream (the direction dictated by shear) to upstream; this is the `population splitting' phenomenon that we thoroughly discussed in our previous work \cite{popsplit-paper}. Increasing angular speed from $\Gamma=0$ to $\Gamma=50$ (levogyre chirality) is seen to lead to a suppression of the minority (and increase of the majority) population peak, hence a lowering of bimodal ratio (defined here as the ratio of the minority over the majority peak population) from $R_{bm}\simeq1/3$ to $R_{bm}\simeq1/11$. 

Staying with levogyre chirality, increasing the angular speed further, to $\Gamma=100$, is shown to lead to a complete suppression of population splitting. In fact, as is seen to be the case for dextrogyre chiral swimmers of the same angular speed, i.e., $\Gamma=-100$, the distribution of active particles becomes close to uniform, or at least more `even', across all possible swimmer orientations. We showed in our previous work that when the strength of imposed shear surpasses a certain threshold, an active suspension of (non-chiral) swimmers will undergo transition from a unimodal to a bimodal regime (distribution). Figure \ref{distribution} shows that with imposed shear unchanged, changing the angular speed, alone, of chiral swimmers may also lead to transitions between unimodal (UM) and bimodal (BM) phases. In fact, Fig. \ref{distribution} suggests that increasing signed angular speed of spheroidal chiral particles from $-100$ to $100$ leads to two transitions: One UM-to-BM transition, followed by a BM-to-UM transition.     
     
\subsection{Effect of chirality on swimmer populations}

\begin{figure}
	\centering
	\includegraphics[width=0.95\linewidth]{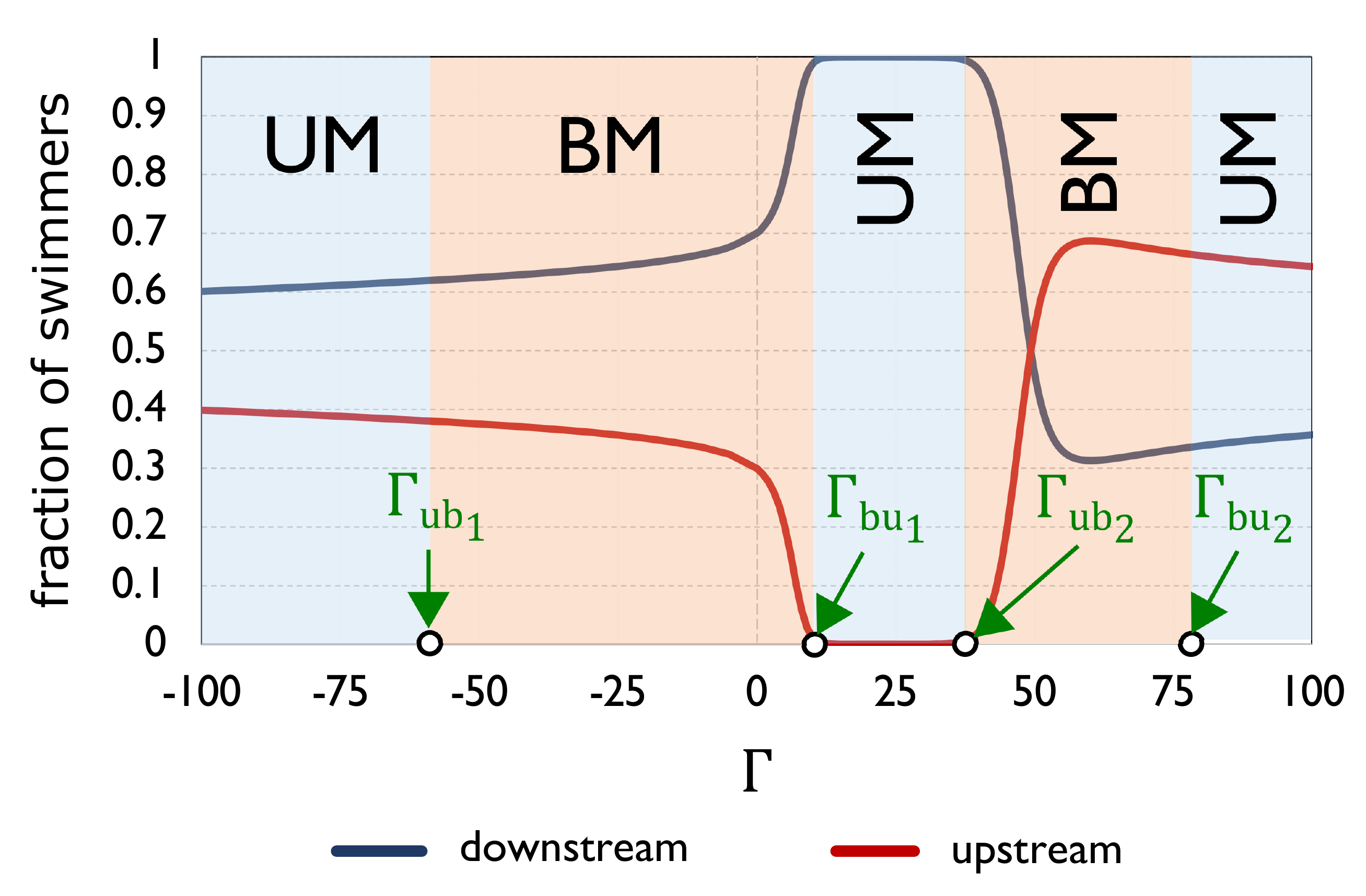}
	\caption{Fractions of the total swimmer population swimming downstream and upstream, at different angular speeds of the levogyre/dextrogyre chiral swimmers. Boxes of two different colors correspond to the two phases of the system: Unimodal (UM) and bimodal (BM). The chiral swimmers are spheroidal with aspect ratio $\lambda=4$ (major and minor axes of lengths $a=2\,\mu {\mathrm{m}}$ and $b=0.5\,\mu {\mathrm{m}}$, respectively).}
	\label{populations}
\end{figure}

By continuous variation of the angular speed, Fig. \ref{populations} provides a closer look into the transitions that an active suspension of chiral swimmers goes through as angular speed $\Gamma$ is varied over a wide range: From dextrogyre chirality with rapid rotation ($\Gamma=-100$) to the non-chiral scenario, and from there to levogyre chirality with rapid rotation ($\Gamma=100$). The plot shows how downstream- and upstream-swimming populations of chiral active particles (as fractions of total swimmer population) vary with angular speed $\Gamma$, with everything else remaining intact, as per our baseline set of parameter values mentioned earlier. We have used boxes of two different colors to represent the two (UM and BM) phases. An immediate observation is that the system goes through four (rather than two) transitions as $\Gamma$ is varied over the range. Starting from $\Gamma=-100$ towards $\Gamma=100$, there is a first UM-to-BM transition (at $\Gamma_{ub_{1}}$), then a BM-to-UM transition (at $\Gamma_{bu_{1}}$), followed by a second UM-to-BM transition (at $\Gamma_{ub_{2}}$), and at last a second BM-to-UM transition (at $\Gamma_{bu_{2}}$). In statistical terms, the bimodality occurring at $\Gamma_{ub_{2}}$ is \textit{re-entrant}.

We start discussing the results illustrated in Fig. \ref{populations} by looking at the vertical midline that represents the non-chiral ($\Gamma=0$) scenario. As was shown earlier (Fig. \ref{distribution}), with non-chiral swimmers, the imposed shear has sufficient strength to cause population splitting; the swimmer population is split into about 70\% swimming downstream, and the remaining 30\% swimming upstream: The regime is bimodal (BM). Dextrogyre chirality works in concert with the imposed flow (see Fig. \ref{channel}), and larger negative (CW) angular speeds enhance the population splitting of swimmers; hence the decreasing downstream (majority) and increasing upstream (minority) populations with increasing angular speed of dextrogyre chiral swimmers. However, beyond $\Gamma_{ub_{1}}$ (i.e., for CW rotation faster than a certain rate), the system is seen to enter the unimodal phase. The presence of population splitting and the unimodal distribution of swimmers might seem contradictory, yet the population and population \textit{peak} distinction clears the potential ambiguity. Populations represent the fraction of active particles swimming down- or upstream, and while there can be two oppositely swimming sub-populations, the swimmer distribution can be unimodal, as there could be no visible \textit{peaks} of a minority population, as was shown to be the case earlier in Fig. \ref{distribution}, where for $\Gamma=-100$ and $\Gamma=100$, the swimmers were seen to be more evenly distributed compared to those with smaller angular speeds. At CW angular speeds larger (in magnitude) than $\Gamma_{ub_{1}}$, the rotation of the chiral swimmers is so fast, and the whole ($0,2\pi$) range of orientation angles is spanned at such rapid rate, that the smaller of the two peaks (i.e., the minority population \textit{peak}) recedes, giving rise to a unimodal distribution (while the \textit{minority} population still exists). For the exact same reason, there is a transition from unimodality to bimodality at levogyre angular speeds larger than $\Gamma_{bu_{2}}$. 

The other two transitions are different, in that they correspond to actual onsets of population splitting. As, on the positive horizontal axis in Fig. \ref{populations}, we move from the non-chiral situation towards positive (CCW) angular speeds for levogyre chiral swimmers, the downstream population is seen to (sharply) increase, to the point that all chiral active particles are swimming downstream, and there is no minority population. This occurs at $\Gamma_{bu_{1}}$, marking the first of two bimodal-to-unimodal transitions for levogyre chiral swimmers. As illustrated in Fig. \ref{schematic}, positive (CCW) angular speed opposes the effect of the imposed Couette flow in exerting a CW torque on the spheroidal swimmers. The angular speed $\Gamma_{bu_{1}}$ is the maximum opposition the imposed flow can bear before its effect in splitting the swimmers into downstream (majority) and upstream (minority) populations is totally cancelled out by that of levogyre chirality. As indicated by the middle UM box in Fig. \ref{populations}, the active suspension remains in the unimodal phase for increasingly large CCW angular speeds, yet at $\Gamma_{ub_{2}}$, there is re-entrant bimodality, i.e., there is transition from the unimodal to the bimodal phase for the second time. For CCW angular speeds larger than $\Gamma_{ub_{2}}$, the downstream population starts decreasing and the upstream population increasing, and therefore the re-entrant bimodality coincides with the onset of a second population splitting. While the first population splitting was initiated by the imposed \textit{flow} attaining sufficient strength to flip the swimming direction of a sizeable fraction of swimmers, this second population splitting is \textit{chirality}-induced. At $\Gamma=\Gamma_{ub_{2}}$, the CCW angular speed of the chiral swimmers has reached sufficient magnitude to give rise to a population splitting of its own: Having overcome the counteracting effect of imposed shear, the CCW torque has become sufficiently strong to flip the orientation of some of the swimmers from downstream to upstream. As schematically shown in Fig. \ref{torques}, the flipping of swimming direction from downstream to upstream can occur under the dominating effect of CW torque (shear-induced), or the dominating effect of CCW torque (chirality-induced). While acting in opposing directions, both shear and chirality can give rise to the conversion of a fraction of downstream-swimming particles to upstream-swimming particles. Figure \ref{populations} also shows that as angular speed of levogyre chiral swimmers is increased beyond the point of transition to re-entrant bimodality ($\Gamma=\Gamma_{ub_{2}}$), the downstream-swimming population decreases, until at some angular speed it becomes equal to the upstream-swimming population. Beyond this point, i.e., for yet larger angular speeds of the levogyre chiral swimmers, the majority and minority populations change places (i.e., exchange orientations), with the upstream-swimming population now forming the majority.

%%%FIG%%%%%%%%%%%%%%%%%%%%%
\floatsetup[figure]{style=plain,subcapbesideposition=top}
\begin{figure}[t!]
\centering
\sidesubfloat[]{%
\centering
  \label{phdiag_v2_a}
  \includegraphics[width=0.925\linewidth]{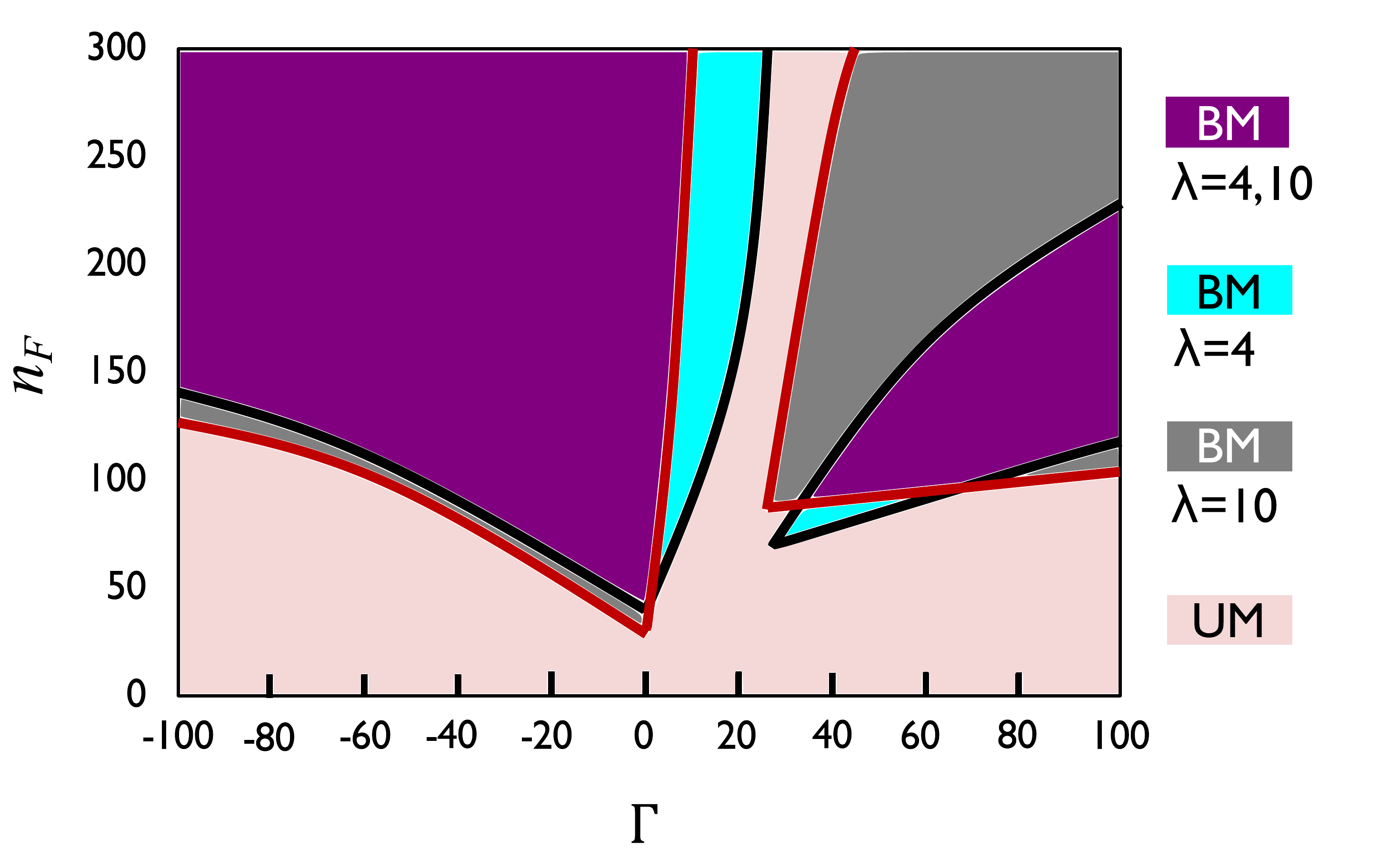}%
}\\
\sidesubfloat[]{%
\centering
  \label{phdiag_v2_b}
\hspace{-4mm}  \includegraphics[width=0.925\linewidth]{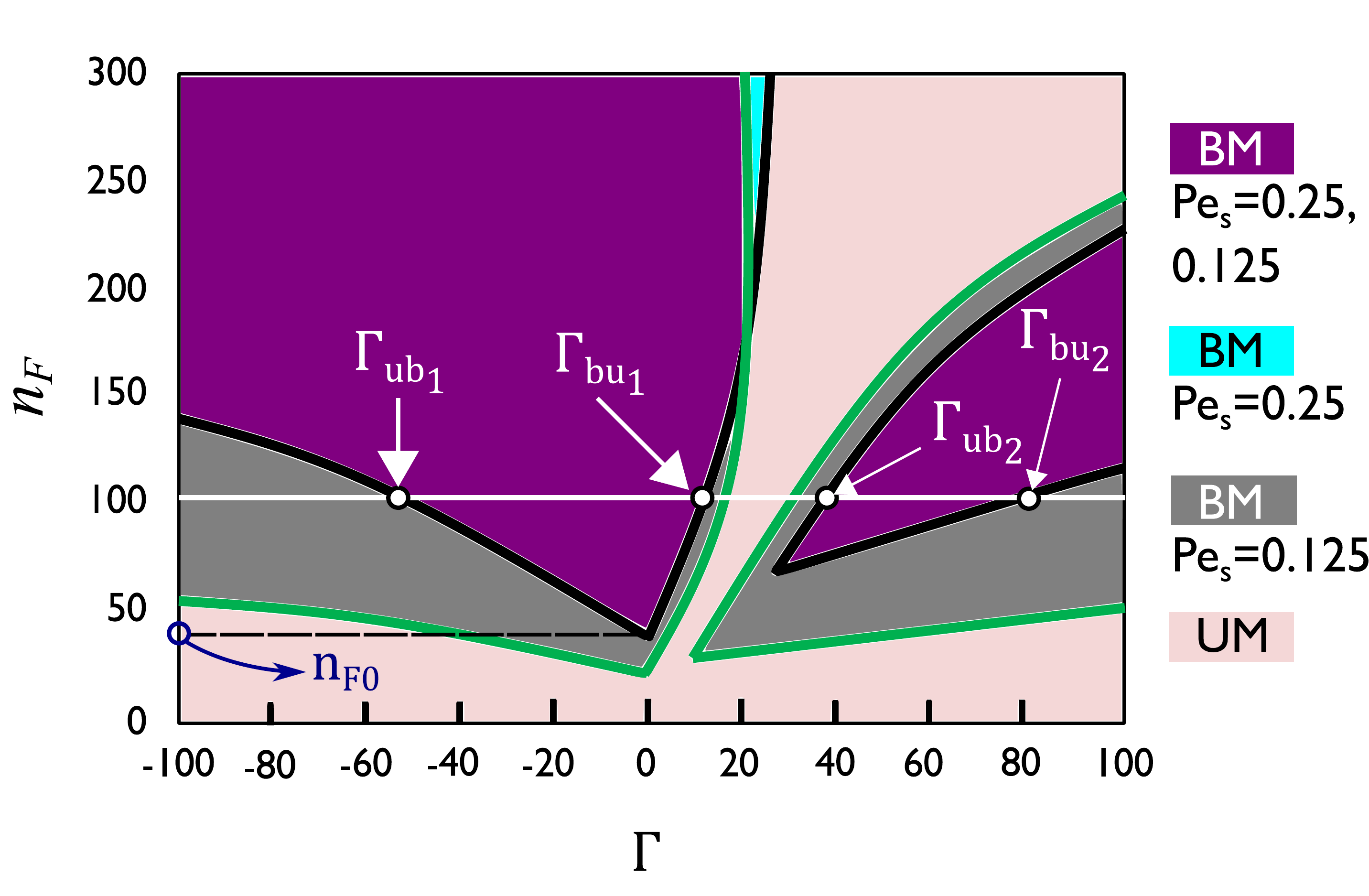}%
  }
\caption{Phase diagrams showing the transitions of spheroidal chiral  swimmers between unimodal (UM) and bimodal (BM) phases as flow factor $n_F$ (representing imposed flow strength) and angular speed $\Gamma$ (due to chirality) are varied along vertical and horizontal axes of the phase diagrams, respectively; (a) phase diagrams for two different particle aspect ratios: $\lambda=4$ (the reference situation) and $\lambda=10$; (b) phase diagrams for two different swimmer self-propulsion strengths, represented by swim P\'eclet numbers $Pe_s=0.25$ and $Pe_s=0.125$ (corresponding to swim speeds $V_s=2\,\mu {\mathrm{m}}/{\mathrm{s}}$, the reference value, and $V_s=1\,\mu {\mathrm{m}}/{\mathrm{s}}$, respectively).
}
\label{phdiags}
\end{figure}
%%%%%%%%%%%%%%%%%%%%%%%%%

\subsection{Phase diagrams}

The data presented in Fig. \ref{populations} is obtained with a given imposed Couette flow, characterized by flow P\'eclet number $Pe_f=25$ or flow factor $n_F=100$ (corresponding to shear rate $\dot{\gamma}=10\,{\mathrm{s}}^{-1}$), from our baseline set of parameter values. The repeated transitions between the two phases of the system (UM and BM) were shown to occur as a result of competition between the torques due to shear and chirality. As shear rate and chirality are crucial in determining system behavior, we present phase diagrams in Fig. \ref{phdiags} that have flow factor $n_{F}$ (representing shear rate) on the vertical and angular speed $\Gamma$ of the chiral swimmers on the horizontal axis. The baseline situation is shown in both Figs. \ref{phdiag_v2_a} and \ref{phdiag_v2_b} (using black lines) for comparison: It shows the four transitions that can occur for a given strength of imposed flow as swimmer chirality is varied over the $\left[ -150,150\right]$ range. It also shows that the number of transitions will depend on the imposed shear rate. At shear rates (or flow factors $n_F$; we shall use the two related parameters interchangeably in our qualitative discussions) smaller than that required to initiate population splitting of \textit{non}-chiral swimmers, chiral swimmers will not experience population splitting either, regardless of the angular speed sign or magnitude; there are no transitions in this range of imposed shear rate. At imposed flows stronger than this threshold (we have shown the threshold flow factor in Fig. \ref{phdiag_v2_b} as $n_{F0}$), there will be two or four transitions between UM and BM regimes, depending on how large the shear rate is. The data in Figs. \ref{distribution} and \ref{populations} pertained to $n_{F}=100$, at which (as can be also seen from Fig. \ref{phdiags}) there are four transitions, with the factors contributing to each of the four discussed earlier above. It can be seen from Fig. \ref{phdiags} that the angular speeds for all four transitions are larger (in magnitude) at larger shear rates. For the two transitions at largest angular speeds ($\Gamma_{ub_{1}}$ and $\Gamma_{bu_{2}}$), it is rapid spanning of the whole $[0,2\pi)$ range of orientation angles that suppresses the effect of imposed shear in giving the active particles a preferred swimming direction. It is therefore according to expectation that stronger imposed flow should face faster rotation (larger magnitude of angular speed) for chirality to dominate and lead to transition. The BM-to-UM transition at $\Gamma_{bu_{1}}$ also occurs at larger angular speeds for stronger imposed flow; as the transition angular speed is the maximum an imposed flow can stand before it loses (effective) strength for initiation of population splitting: Stronger imposed flow can stand larger angular speeds due to chirality. The UM-to-BM transition at $\Gamma_{ub_{2}}$ occurs when CCW rotation due to levogyre chirality overcomes the effect of imposed shear and gives rise to a population splitting of its own. Stronger flow would have to be overcome by larger (CCW) angular speeds, i.e., more `power' from chirality. 

A less trivial observation from Fig. \ref{phdiags} is that for a range of imposed shear rates, all larger than that required to initiate population splitting of non-chiral swimmers, only two transitions occur. The observation implies that for the second population splitting to take place (at $\Gamma_{ub_{2}}$), even though it is driven by chirality, the imposed shear rate needs to be greater than a threshold. Chirality of the active particles, independently, and without the presence of imposed shear beyond a certain strength, cannot give rise to population splitting of the swimmers. This is expected to be true for shear-driven population splitting; Fig. \ref{phdiags} shows that this is true also for chirality-driven population splitting of the swimmers. At weaker imposed flow, the angular speed $\Gamma_{bu_{1}}$, at which the first BM-to-UM transition takes place, is the point beyond which the swimmer distribution starts verging towards increased evenness (eventually taking the shape of a nearly uniform distribution), so that, in effect, it coincides with the angular speed $\Gamma_{bu_{2}}$, never giving chance for the rising of a minority population peak. 

\subsection{Effect of swimmer aspect ratio}

Figure \ref{phdiag_v2_a} shows the effect of swimmer aspect ratio on the behavior of an active suspension of chiral swimmers subject to imposed shear. It can be seen that particle aspect ratio mostly affects the transitions specific to levogyre chiral swimmers, occurring at $\Gamma_{bu_{1}}$ and $\Gamma_{ub_{2}}$: In both cases, the angular speed (due to chirality) at which the transition occurs is smaller (at a given imposed shear rate) for thinner (larger aspect ratio) swimmers. This can be explained by the larger rotational diffusivity of thinner particles. As the effect of the imposed shear is to orient the chiral swimmers along the direction of flow, i.e., horizontally (in or against the flow), increased rotational diffusion of thinner swimmers is a hindrance to this task, in that larger $D_R$ would imply larger resistance against remaining in a certain direction. With the effect of the imposed flow in bringing about population splitting (by aligning the chiral swimmers horizontally) is weakened for thinner chiral particles, the imposed flow will lose its ability to maintain the population splitting marked by $\Gamma_{bu_{1}}$ at an angular speed smaller than that for a swimmer of smaller aspect ratio, hence the smaller $\Gamma_{bu_{1}}$. The angular speed $\Gamma_{ub_{2}}$ marking the onset of chirality-driven population splitting is also smaller for thinner particles due more rotational diffusion of the chiral swimmers making it easier for CCW chiral torque to overcome the effect of imposed shear in horizontally aligning the active spheroidal particles. 

\subsection{Effect of swimmer self-propulsion strength}
The effect of self-propulsion speed on the behavior of the confined active suspension of spheroidal chiral swimmers can be seen from the phase diagram of Fig. \ref{phdiag_v2_b}. In contrast with swimmer aspect ratio, self-propulsion speed is seen to affect the transitions at largest angular speeds much more than the other two transitions. This can be explained by the fact that the latter two transitions arise from a competition of imposed shear and chirality, with stronger or weaker active self-propulsion not having a major say. But the transitions to unimodal distribution at large magnitudes of angular speed (due to chirality) occur when all orientation angles are spanned at a very rapid rate, leading to even distribution of swimmers across all $\theta$. Stronger swimmer self-propulsion works to have the chiral particles oriented vertically toward channel walls, and in doing so resists the action of large angular speeds at spanning the whole circle of radiation at very rapid rates, leading to nearly uniform distributions.   

\section{Conclusions}
We presented quantitative analysis on the behavior of a dilute active suspension of spheroidal chiral  swimmers, in confinement, subjected to imposed shear. Having shown in previous work \cite{popsplit-paper} that imposed flow beyond a certain strength gives rise to the splitting of swimmers into distinct downstream and upstream populations, we showed here that for chiral swimmers, the picture is considerably more nuanced, with the occurrence of population splitting (as characterizer of the response of an active suspension to shear) showing strong dependency on swimmer chirality: Angular speed and direction of rotation (levogyre/dextrogyre chirality). We attributed two phases to the system, corresponding to the presence of one or two peaks in the swimmer distribution function across all orientation angles; namely, unimodal and bimodal phases, respectively. Using phase diagrams covering a wide range of chiralities and imposed shear rates, we showed that the active suspension could switch states (transit between phases) upon modest changes in the angular speed of the swimmers and/or the shear rate of imposed flow. Considering variations of chiral swimmer angular speed at a given imposed shear rate, we observed re-entrant bimodality in the active suspension, meaning that under otherwise identical circumstances, chiral swimmers with a given angular speed $\Omega_1$ could be in the bimodal phase, while those with larger angular speeds $\Omega_2$ could be in the unimodal \textit{or} bimodal phase depending on how larger $\Omega_2$ is, compared to $\Omega_1$. Considering increasing $\Omega_{2}-\Omega_{1}$ from $0$ to larger values, the chiral swimmers will be in the bimodal, unimodal, bimodal (again) and unimodal (again) phases as the difference in angular speeds is gradually increased. We further showed, based again on phase diagrams, that the state of the active suspension is notably different for chiral swimmers of different aspect ratios (albeit propelling at the same speed and in the same direction, and subject to the same imposed shear rate). We also showed that otherwise identical swimmers (subject to the same shear rate) may or may not flip their swimming direction to the exact opposite (of that dictated by the imposed flow) depending on their self-propulsion speed. The observations suggest the possibility of using imposed shear as control factor to sort the chiral swimmers in an active suspension according to particle aspect ratio, self-propulsion, or angular speed. With the three mentioned features characterizing swimmers of different types, applications could be envisaged for sorting/separating biological or artificial self-propelled particles of different specifications.

\begin{acknowledgments}
A.N. acknowledges partial support from Iran Science Elites Federation (ISEF) and the Associateship Scheme of The Abdus Salam International Centre for Theoretical Physics (Trieste, Italy). We thank M. Kheyri and M.R. Shabanniya for  useful discussions. 
\end{acknowledgments}

\bibliography{chiralityrefs}

\end{document}